\documentclass[twocolumn,prb,showpacs,amsfonts,amsmath,amssymb,superscriptaddress]{revtex4-1}

\usepackage{graphicx}
\usepackage{amsfonts}
\usepackage[latin1]{inputenc} 

\newcommand{\ud}{\mathrm{d}}
\newcommand{\nn}{\nonumber}

\newcommand{\aou}{\hat{a}}
\newcommand{\aod}{\hat{a}^{\dagger}}
\newcommand{\ai}{\hat{a}}
\newcommand{\aid}{\hat{a}^{\dagger}}
\newcommand{\n}{\hat{n}}
\renewcommand{\ni}{\n}

\newcommand{\bt}[1]{\langle#1\rangle}

\makeatletter

\newcommand{\Rmnum}[1]{\expandafter\@slowromancap\romannumeral #1@}
\makeatother

\begin{document}

\title{Continuous-wave spatial quantum correlations of light induced by multiple scattering}

\author{Stephan Smolka}
\email{smolka@phys.ethz.ch}
\thanks{Present address: Institute of Quantum Electronics, ETH Z\"urich, 8093 Z\"urich, Switzerland}
\affiliation{DTU Fotonik, Department of Photonics Engineering, Technical University of Denmark, Building 345V, 2800
Kgs. Lyngby, Denmark}

\author{Johan R. Ott}
\affiliation{DTU Fotonik, Department of Photonics Engineering, Technical University of Denmark, Building 345V, 2800
Kgs. Lyngby, Denmark}

\author{Alexander Huck}
\affiliation{DTU Physics, Department of Physics, Technical University of Denmark, Building 309, 2800 Kgs. Lyngby, Denmark}

\author{Ulrik L.\ Andersen}
\affiliation{DTU Physics, Department of Physics, Technical University of Denmark, Building 309, 2800 Kgs. Lyngby, Denmark}

\author{Peter Lodahl}
\email{lodahl@nbi.ku.dk}
\homepage{\\ www.quantum-photonics.dk}
\affiliation{ Niels Bohr Institute, University of Copenhagen, Blegdamsvej 17, Dk-2100 Copenhagen, Denmark}

\begin{abstract}
We present theoretical and experimental results on spatial quantum correlations induced by multiple scattering of nonclassical light. A continuous mode quantum theory is derived that enables determining the spatial quantum correlation function from the fluctuations of the  total transmittance and reflectance. Utilizing frequency-resolved quantum noise measurements, we observe that the strength of the spatial quantum correlation function can be controlled by changing the quantum state of an incident bright squeezed-light source. Our results are found to be in excellent agreement with the developed theory and form a basis for future research on, e.g., quantum interference of multiple quantum states in a multiple scattering medium.
\end{abstract}


\maketitle


\section{Introduction}

The light transport through an inhomogeneous random medium is determined by a multiple scattering process whereby interference between different light paths induces a complex intensity speckle pattern.  Mesoscopic intensity fluctuations that survive averaging over all configurations of disorder give rise to phenomena such as enhanced backscattering  and Anderson localization of light.~\cite{sheng} In order to characterize such a disordered medium, it is essential to investigate spectral or temporal intensity correlations in the multiply scattered light.~\cite{prl61p459,prl61p834,revmodphys71p313} In the diffusive regime, the light intensities of different spatial directions (i.e. independent speckles) are uncorrelated. In this case interference effects disappear after ensemble averaging and the transport is well described by diffusion theory.

While the classical aspects of multiple scattering have been studied extensively, the quantum nature of light in a multiple scattering setting has been addressed only recently. Triggered by theoretical studies on quantum fluctuations of light in absorbing and amplifying multiple scattering media,~\cite{prl81p1829,physreva61p063805} recent experiments have addressed the quantum noise of multiple scattered light~\cite{prl94p153905,prl97p103901,optlett31p110} and spatial photon correlations.~\cite{prl95p173901,prl102p193901,prl104p173601,arxiv10004_1721} Furthermore, first studies of photon emission and cavity quantum electrodynamics in disordered media have been carried out~\cite{science327p1352, prl105p013904,njp13p063044,prl108p113901}.
  It was recently predicted that quantum interference and entanglement can be induced by combining two or more quantum states in a multiple scattering medium,~\cite{prl102p193601,prl105p090501} which could be of potential interest for applications in quantum information processing. In the diffusive regime of multiple scattering classical intensity correlations vanish and many quantum optical phenomena do not survive ensemble averaging including polarization entanglement~\cite{physreva70p023808} and quadrature squeezing.~\cite{optexpress14p6919} In contrast, it has been proposed that nonclassical photon fluctuations can be detected in the multiple scattered light even after ensemble averaging, which results in a new type of correlations that is of quantum origin.~\cite{prl95p173901} These spatial quantum correlations have been reported recently based on total transmission quantum noise measurements, thereby entering for the first time experimentally the genuine quantum optical regime of multiple scattering.~\cite{prl102p193901}

In this report, we present a detailed theoretical and experimental analysis of spatial quantum correlations in transmission and reflection measurements. We develop in Sec.~\Rmnum{2} a continuous mode quantum theory in order to relate the spatial quantum correlation function to the photon fluctuations of the light source. By exploiting quantum noise measurements, the variance of the photon fluctuations of the multiple scattered light is studied in Sec.~\Rmnum{3} from which the strength of the spatial quantum correlation function is determined. Finally, it is experimentally shown how the spatial quantum correlation is tuned by controlling the quantum state of the light source.

\section{Theory}
\begin{figure}[ht]
  \centering
  \includegraphics[width=0.3\textwidth]{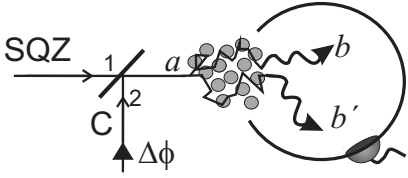}
\caption{Illustration of multiple scattering leading to spatial quantum correlations. The nonclassical light source that is used in the experiment is created by overlapping a squeezed vacuum light beam (SQZ) and a coherent light beam (C) on a beam splitter with two input ports, 1 and 2, respectively. The relative phase, $\Delta \phi$, between C and SQZ can be tuned continuously. The resultant light beam in direction $a$ is incident onto a medium with randomly distributed scatterers. Light is split into a multitude of different trajectories and the number of photons exiting the medium in a specific direction, $b$, can be correlated with the number of photons in another direction, $b'$, to a degree dependent on the quantum state of the illuminating light source. In the diffusive regime the spatial quantum correlation function can be determined from total transmission quantum noise measurements. A similar experimental scheme is applied for reflection measurements.}
  \label{fig01}
\end{figure}

The quantum nature of light is revealed, e.g., by investigating photon fluctuations and spatial, temporal, or spectral quantum correlations among photons. In the experiment of concern here, the quantum properties are identified through the photon fluctuations for which sub-Poissonian fluctuations is a sign of non-classicality.~\cite{Loudon} A schematic illustration of the experimental setup  is displayed in Fig.~\ref{fig01}: a bright squeezed state of light is sent through a multiple scattering medium, the total transmitted or reflected light is collected, and the associated photon fluctuations are recorded. In the following, we introduce a continuous mode quantum theory that is directly applicable to the quantum noise measurements discussed in the present work.

We define the spectral density function at frequency $\omega$ as the Fourier transformation of the autocorrelation function~\cite{madelwolf}
\begin{align}
S(\omega)=&\frac{1}{2\pi}\int_{-\infty}^{\infty}\ud \tau\Gamma(\tau)e^{i\omega \tau}, \label{eq_noisepower}
\end{align}
where
\begin{align}
{\Gamma(\tau)}=& {\bt{\ni(t)\ni(t+\tau)}}-{\bt{\ni(t)}\bt{\ni(t+\tau)}},
 \end{align}
 with $\ni(t)$ being the operator describing the photon flux.  The variance of the photon fluctuations is related to the noise spectrum through the inverse of Eq.~\eqref{eq_noisepower}, i.e.  $\Delta \ni^2(t)=\int_{-\infty}^{\infty}\ud \omega S(\omega)$. The spectral density  is studied within a frequency region $\omega_\pm =\omega_{s}\pm\delta\omega/2$
\begin{align}
\left<\Delta n^2(\omega_s,\delta \omega)\right>=\int_{\omega_-}^{\omega_+} \ud\omega\, S(\omega)\label{eq_noisepower2}
\end{align}

that is proportional to the experimentally measured noise power.

The randomly disordered medium is treated as $N$ spatially distinct optical input modes that are intrisically connected to $N$ output modes via multiple scattering of light. In order to relate the quantum properties of the multiple scattered light to the quantum state of the incident light source, the spatial output mode $b$ is related to all input modes $a'$ through
\begin{align}\label{SQC:eq03}
\aou_b(t)=\sum_{a'} s_{a'b}\ai_{a'}(t),
\end{align}
where $\ai_{b}(t)$ denotes the time-dependent annihilation operator and $s_{a'b}$ is the complex electric field amplitude scattering coefficient corresponding to either a reflection $(r_{ab})$ or transmission $(t_{ab})$  channel. The corresponding intensity transmission (reflection) coefficient is $T_{ab}\equiv|t_{ab}|^2$ ($R_{ab}=|r_{ab}|^2$). The annihilation and creation operators obey the commutation relation $[\aou_b(t),\aod_{b'}(t')]=\delta_{b,b'}\,\delta(t-t')$, with $\delta(t-t')$ and $\delta_{b,b'}$ being the Dirac and Kronecker delta functions, respectively. The average photon flux is defined as $\bt{\ni_b(t)} = \bt{\aod_b(t)\aou_b(t)}$ while the variance in the photon fluctuations is $\Delta n_b^2(t)=\bt{\ni_b^2(t)}-\bt{\ni_b(t)}^2$, where $\bt{\cdots}$ describes the quantum mechanical expectation value. In the following we focus on the setting where the light beam is incident onto the medium through a single direction, $a$, thus the average photon flux for all other input modes equals zero. The mean photon flux of the total transmission can be obtained by summing over all transmitted output modes, i.e. $\bt{\ni_T(t)}=\sum_{b}T_{ab}\bt{\aid_{a}(t)\ai_{a}(t)}$, where the intensity transmission coefficients are random variables determined by the individual realization of disorder. Predictable physical variables are obtained after ensemble-averaging over all configurations of disorder, e.g., the total transmitted photon flux is given by
\begin{align}
\overline{\bt{\ni_T(t)}}=&\overline{T} \bt{\ni(t)},\label{cross_corr}
\end{align}
where $T=\sum_{b}T_{ab}$ is the total intensity transmission for the flux of photons $\bt{\ni(t)}$  incident in mode $a$, and the bar denotes the ensemble average over all configurations of disorder.~\cite{revmodphys71p313} A similar result can be found for the total reflection, i.e. $\overline{\bt{\ni_R(t)}}=\overline{R} \bt{\ni(t)}$ where $R=\sum_bR_{ab}$.
In order to determine the fluctuations of the total transmission we need to evaluate the second moment of the photon number operator
\begin{align}
\overline{\bt{\ni_T(t)\ni_T(t+\tau)}}=& \sum_{b, b'}\overline{\bt{\ni_{b}(t)\ni_{b'}(t+\tau)}}\nn\\
=& \overline{T^2}\bt{\ni(t)\ni(t+\tau)}+ (\overline{T}-\overline{T^2})\bt{\ni(t)}\delta(\tau).\label{cross_corr2}
\end{align}
The non-vanishing contributions to the sum in Eq.~\eqref{cross_corr2} for different output modes $b$ and $b'$ stem from the fact that different spatial parts of the volume speckle patterns are quantum correlated.
In an experiment the total collection and detection efficiency is non-unity, which is accounted for by substituting $\overline T \rightarrow \eta_{T}\overline T$. The photon fluctuations of the total transmission, $\left<\Delta n^2_{T}(\omega_s,\delta \omega)\right>$, can be related to the spectral density function of the incident light source integrated over the measured frequency range from Eqs. (\ref{eq_noisepower}-\ref{eq_noisepower2})
\begin{align}\label{gamma_freq_T}
\nonumber &\overline{\bt{\Delta n^2_{T}(\omega_s,\delta \omega)}}\\
\nn =&\int_{\omega_-}^{\omega_+} \ud \omega \,\left[\eta^2\overline{T}^2\left(S(\omega) - \frac{\bt{\ni(t)}}{2 \pi}\right)+\eta\overline T \frac{\bt{\ni(t)}}{2 \pi}\right],\\
=& \eta^2\overline{T}^2\left(\bt{\Delta n^2(\omega_s,\delta \omega)} - \bt{\ni(t)}\,\frac{\delta \omega}{2 \pi}\right)+\eta\overline{T}\bt{\ni(t)}\,\frac{\delta \omega}{2 \pi},
\end{align}
where we have used $\overline{T^2}=\overline{T}^2$ that holds in the diffusive regime.~\cite{revmodphys71p313} Evaluating the last expression for a coherent state, yields $\overline{\bt{\Delta n^2_{C,T}(\omega_s,\delta \omega)}}=\eta\overline{T}\bt{\ni_C(t)}\,\delta \omega/2 \pi$.

We introduce the Fano factor
\begin{align}\label{eq:Fano_factor}
F(\omega,\delta\omega)\equiv&\frac{\bt{\Delta n^2(\omega,\delta\omega)}}{\bt{\Delta n^2_{C}(\omega,\delta\omega)}},
\end{align}
that gauges the ratio between the variance in the photon fluctuations of an unknown quantum state, $\left<\Delta n^2(\omega_s,\delta \omega)\right>$, and of a coherent state, $\left<\Delta n^2_C(\omega_s,\delta \omega)\right>$ having the same mean photon flux. Importantly, the Fano factor can be directly measured without determining the proportionality factor between the noise power recorded with the electrical spectrum analyzer and the photon number variance. In the experiment the optical state is created by overlapping a squeezed-vacuum state with a bright-coherent state, cf. Fig.~\ref{fig01}. Since the resultant bright-squeezed state has a large coherent amplitude, the average photon flux is to a very good approximation equal to the average photon flux of the coherent state and Eq.~\eqref{gamma_freq_T}  can be rewritten as
\begin{align}
\nn \overline{F_T}(\omega_s,\delta \omega)=&\frac{\eta_{T}^2\overline{T}^2\left(\bt{\Delta n^2(\omega_s,\delta \omega)} - \bt{\ni(t)}\,\delta \omega/2 \pi\right)}{\eta_{T}\overline{T}\bt{\ni(t)}\,\delta \omega/2 \pi}+1\\
=&1-\eta_{T}\overline{T} \left[1- F(\omega_s,\delta \omega)\right]. \label{eqns:Fano}
\end{align}
A similar expression holds for the total reflection, i.e., $\overline{F_R}(\omega_s,\delta \omega)=1-\eta_{R}\overline{R} \left[1- F(\omega_s,\delta \omega)\right]$ with $\eta_{R}$ being the corresponding detection efficiency. We emphasize the importance of the last expressions. In Ref.~[\onlinecite{optexpress14p6919}] it was predicted that any nonclassical features in the electric field quadrature amplitudes vanish in the multiple scattering process after ensemble averaging. In contrast, Eq.~\eqref{eqns:Fano} shows that for a quantum state with nonclassical photon fluctuations also the ensemble-averaged transmitted and reflected photon fluctuations exhibit nonclassical photon fluctuations after multiple scattering.

We now introduce the spatial quantum correlation function, which is the quantity extracted in the experiment. It is defined as
\begin{align}
 \overline{C_{bb'}^{Q}}(\omega_s,\delta\omega) = \frac{1}{2 \pi} \int_{-\infty}^{\infty} d \tau \int_{\omega_-}^{\omega_+} d \omega \overline{C_{bb'}}(\tau)\,e^{i\omega\tau},
\end{align}
for $b \neq b'$ where
  \begin{align}
\overline{C_{bb'}}(\tau)=\frac{\overline{\bt{\ni_b(t)\ni_{b'}(t+\tau)}}}{\overline{\bt{\ni_b(t)}}\times\overline{\bt{\ni_{b'}(t+\tau)}}}.\label{eqn:corr}
\end{align}
The correlation function gauges the photon correlations between two different output modes $b$ and $b'$. Starting from Eq. (\ref{SQC:eq03}) we derive
\begin{align}
\nonumber &\overline{C_{bb'}^{Q}}(\omega_s,\delta\omega)\\ 
\nn &= \frac{1}{2 \pi}\int_{-\infty}^{\infty} d \tau\int_{\omega_-}^{\omega^+}\ud\omega\, \frac{\Gamma(\tau)-\bt{\ni(t)}\delta(\tau)+\bt{\ni(t)}^2}{\bt{\ni(t)}^2}\overline{C_{bb'}^C}\,e^{i\omega\tau}, \\
 &= \frac{\left[F(\omega_s,\delta\omega)-1\right]}{\bt{\ni(t)}}\,\frac{\delta\omega}{2\pi}\,\overline{C_{bb'}^C},\label{corr_freq_incident}
\end{align}
by using that $\bt{\ni(t)}\approx\bt{\ni(t)}_C=2\pi/\delta\omega\,\bt{\Delta\ni^2_C(\omega_s,\delta\omega)}$.  Here $\overline{C_{bb'}^C}=\overline{T_{ab}T_{ab'}}/(\overline{T_{ab}}\times\overline{T_{ab'}})$ is the classical speckle correlation function that is induced by multiple scattering~\cite{prl61p459,prl61p834,revmodphys71p313}.

From Eq. (\ref{corr_freq_incident}) it can be seen that different optical modes $b$ and $b'$ are positively correlated, i.e., the photons show spatial anti-bunching ($\overline{C_{bb'}^{Q}}(\omega_s,\delta\omega)\geq0$) when the light source exhibits classical photon fluctuations. For a light source that exhibits nonclassical photon fluctuations ($F(\omega_s,\delta\omega)<1$), we enter the purely quantum optical regime where photons propagating two different directions have negative spatial correlations ($\overline{C_{bb'}^{Q}}(\omega_s,\delta\omega)<0$). The spatial quantum correlation function can be expressed in terms of the ensemble averaged total transmitted (reflected) Fano factor that is directly measured in the experiment. This is seen by substituting Eq.~\eqref{eqns:Fano} that holds in the diffusive regime $(\overline{C_{bb'}^{C}}=1)$ into Eq.~\eqref{corr_freq_incident}, leading to the expression
\begin{align}\label{corr_freq_transmitted}
 \overline{C_{bb'}^{Q}}(\omega_s,\delta\omega) &= \frac{\left[\overline{F_{T/R}}(\omega_s,\delta \omega)-1\right]}{\overline{\bt{\ni_{T/R}(t)}}}\,\frac{\delta\omega}{2\pi}.
\end{align}
Thus, the spatial quantum correlation function can be extracted by recording in addition to the Fano factor also the photon flux and the detection efficiency for the total transmission or reflection from the multiple scattering medium, as schematically shown in Fig.~\ref{fig01}. The mean photon flux can be measured by recording the light power of the light source, $P$, divided by the energy of a photon at the optical frequency, $\omega_0$, such that $\langle \hat n(t) \rangle = P /\hbar\omega_0$, where $\hbar$ is the Planck constant divided by $2\,\pi$. Our method provides an experimentally simple way of extracting two-channel speckle correlations in comparison with a direct measurement based on correlation measurements with two single photon counting detectors.~\cite{arxiv10004_1721}

\section{Experiment}
In the following the performed experiment is discussed. First, the experimental setup and the characteristics of the used light source are described in Sec.~\Rmnum{3}.A. A detailed characterization of the multiple scattering sample is carried out in Sec.~\Rmnum{3}.B. In Sec.~\Rmnum{3}.C we present the experimental results on the measured multiple scattered photon fluctuations and show how the magnitude of the spatial quantum correlations can be modified.

\subsection{The squeezed light source}
\begin{figure}[ht]
  \centering
  \includegraphics[width=0.5\textwidth]{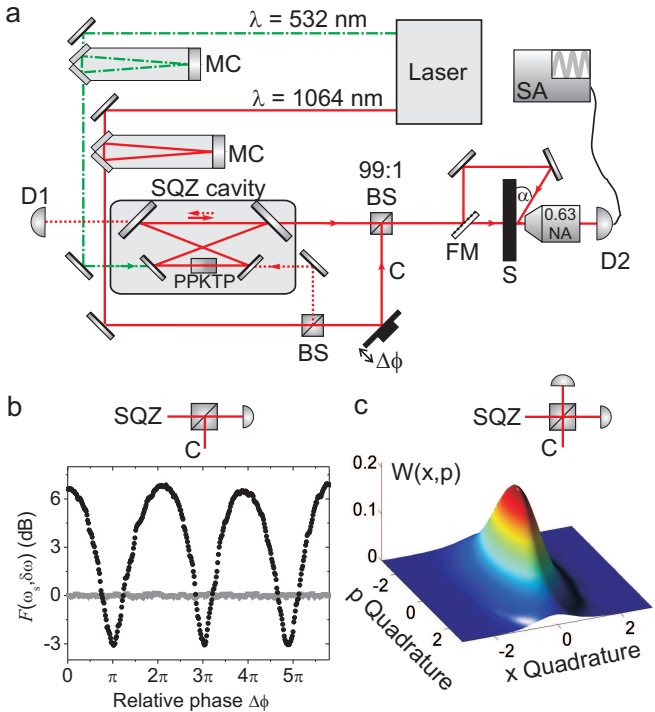}
\caption{(color online). \textbf{a}, Sketch of the experimental setup. MC: mode cleaner, PPKTP: periodically poled potassium titanyl phosphate nonlinear crystal as part of an optical parametric oscillator (gray shaded box), SQZ: vacuum squeezed light beam, C: coherent light beam with a variable optical phase, $\Delta \phi$, relative to the squeezed vacuum, BS: beam splitter, FM: flip mirror, S: multiple scattering sample, SA: spectrum analyzer, D: detector. The detection detector, $D2$, has a bandwidth of $315\,$kHz. The multiple scattered light is collected with a microscope objective featuring a numerical aperture of NA=0.63. In reflection geometry the incident light illuminates the sample under an angle $\alpha=69^\circ$. A detailed explanation of the setup can be found in the main text. \textbf{b}, Measured Fano factor, $F(\omega_s,\delta \omega)$, at an elecrical sideband frequency of $\omega_s=3.93\,$MHz of the bright squeezed light source without insertion of the multiple scattering sample (black circles) depending on the phase $\Delta \phi$ of the coherent beam, as schematically shown in the inset. The measured classical limit is plotted with gray triangles and is obtained by blocking the squeezed beam. As the resolution bandwidth we used $\delta \omega/2\pi = 300\,$kHz. The uncertainty in the Fano factor is estimated over 18 full periods of $\Delta\phi=0\ldots36\pi$.  \textbf{c}, Reconstructed Wigner function of the squeezed vacuum state (SQZ) produced by the optical parametric oscillator. The Wigner function is obtained from time resolved measurements  using a homodyne setup with a 50:50 beam splitter (see inset) and quantum state reconstruction.}
  \label{fig02}
\end{figure}

The experimental setup to study the photon fluctuations of multiply scattered light is schematically shown in Fig.~\ref{fig02}a. The squeezed vacuum state is generated with a bow-tie shaped optical parametric oscillator operating below threshold.~\cite{apl89p061116,optexpress7p4321} As a nonlinear medium, a type~\Rmnum{3} periodically poled nonlinear crystal was placed inside the optical parametric oscillator and pumped with a continuous wave laser (Diabolo, Innolight) at a wavelength of $\lambda_p = 532\,$nm. A second-order nonlinear process generates photon pairs that are in resonance with the cavity and are centered around $\lambda_0 = 1064\,$nm. A fraction of the cavity mode is coupled out by a partly transmitting cavity mirror with a power transmission of $10\,\%$. By operating the optical parametric oscillator below threshold, a squeezed vacuum state is generated.~\cite{prl57p2520} One of the cavity mirrors is mounted on a piezo-electric element in order to adjust the optical path length of the cavity. The optical parametric amplifier is stabilized by recording the transmittance through the cavity of a counter-propagating light beam on detector D1 (red dashed line) using the Pound-Drever-Hall technique that stabilizes the cavity over several hours.~\cite{applphysb31p97} A bright amplitude-squeezed state is generated by overlapping the squeezed vacuum state with a coherent state on a non-polarizing beam splitter that transmits 99\% of the squeezed vacuum state and mixes in $1\,\%$ of the strong coherent state, cf. Fig.~\ref{fig02}a. Note that the relative phase $\Delta \phi$ between the coherent state and vacuum squeezed state must be chosen properly in order to generate a bright amplitude-squeezed state. By continuously tuning the relative phase, $\Delta \phi$, a light source with excess or reduced photon fluctuations relative to the classical limit is obtained. Both the pump beam and the coherent displacement beam are shaped in high finesse modecleaning cavities (MC) in order to ensure a good spatial mode matching of the vacuum squeezed state and the coherent state on the beam splitter.

In order to characterize the bright squeezed light source, we use a highly sensitive InGaAs-resonance detector that converts the photo current of the photo diode (ETX 500T, Epitaxx) with a transimpedance amplifier into a voltage.  An electrical spectrum analyzer computes the Fourier spectrum of the auto correlation function, $\Gamma(\tau)$ (cf. Eq.~\eqref{eq_noisepower}), from the AC voltage of the photo detector at an electrical sideband frequency $\omega_{s}$ with a resolution bandwidth of $\delta \omega/2 \pi=300\,$kHz. The resonance frequency of our photo detector is $3.93\,$MHz, which is therefore the chosen sideband frequency $\omega_{s}/2\pi=3.93\,$MHz in our experiment. Next, we relate the DC voltage of the photo detector to the mean photon flux by calibrating it with a power meter. To this end we measure the light power, $P$ that is proportional to the mean photon flux and divide it by the energy of a single photon, i.e. $\langle \hat n(t) \rangle=P/\hbar\,\omega_0$. The linearity of the photo detector is ensured by measuring the DC and AC voltage of the photodiode versus power of the coherent laser beam. The experiments are performed with optical powers of the light source larger than $P = 5\,\mu$W where the signal is $6.5\,$dB larger than the dark noise level of the detector.

Figure~\ref{fig02}b shows the measured Fano factor, $F(\omega_s,\delta \omega)$, of the light source as a function of the relative phase between the coherent beam and the squeezed vacuum beam, $\Delta \phi$, as obtained by removing the multiple scattering sample. The classical limit (gray triangles) is recorded by blocking the squeezed vacuum beam, thereby probing the bright coherent beam. The relative phase, $\Delta \phi$, is scanned with a mirror mounted on a piezo-electric element. We find that the photon statistics of the nonclassical light source can be continuously tuned below the classical limit $(F(\omega_s,\delta \omega) = 1)$ to a minimum of $F(\omega_s,\delta \omega)=0.52 \pm 0.02$ and above the classical limit to a maximum Fano factor of $F(\omega_s,\delta \omega) = 4.6\pm  0.2$.

We furthermore characterize the vacuum squeezed state by the technique of quantum tomography where the Wigner function is reconstructed based on phase-sensitive measurements of the quantum fluctuations, see inset in Fig.~\ref{fig02}c. Two photo detectors mounted at the output ports of a 50:50 beam splitter record the photo-current as a function of the time. Both time-resolved signals are subtracted from each other, mixed with an electronic local oscillator at a frequency of $3.93\,$MHz, and low-pass filtered with a bandwidth of $150\,$kHz. The relative phase, $\Delta\phi$, of the coherent light beam is slowly varied so that the Wigner function is reconstructed from the recorded data for different values of the phase, using an iterative maximum-likelihood reconstruction algorithm.~\cite{JopB6pS556} The Wigner function is a quasi-probability distribution that contains the complete information about the quantum state. It is expressed in terms of the real and imaginary part of the quantized electric field, the so-called $x$ and $p$ quadratures. From Fig.~\ref{fig02}c we observe that the vacuum squeezed state has strongly reduced fluctuations in the $x$ quadrature at the expense of enhanced fluctuations in the $p$ quadrature fulfilling the Heisenberg uncertainty relation.~\cite{physrevd23p1693}

The multiple scattering experiment is conducted by focusing the nonclassical light onto either the front surface or the back surface of the sample, cf. Fig.~\ref{fig02}a, to perform total transmission and total reflection measurements, respectively. The multiply scattered light is collected with a microscope objective ($NA=0.63$) and recorded with the photo detector $D2$. In the reflection geometry it is desirable to remove direct reflections due to single-scattering events from the multiply scattered distribution of light. To achieve this, the multiple scattering surface of the sample is illuminated under a steep angle of $69^\circ$, whereby the direct reflection does not pass through the collection microscope objective.

\subsection{Sample characterization}

\begin{figure}[ht]
  \centering
  \includegraphics[width=0.45\textwidth]{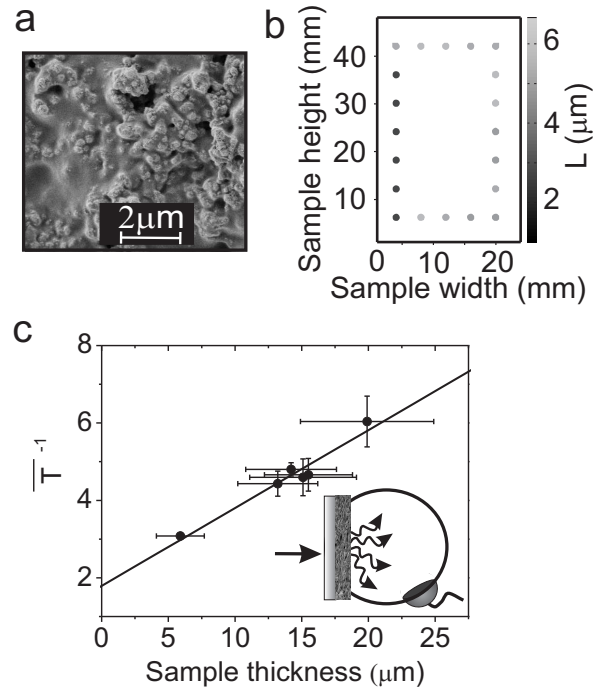}
\caption{\textbf{a}, Scanning electron microscope image of the multiple scattering medium consisting of TiO$_2$ that has been grinded into strongly scattering particles with a typical size of $200\:\text{nm}$ and deposited on a glass substrate. \textbf{b}, Scanning of a sample at different positions with the tip of a micrometer screw to obtain the average sample thickness. The scanned positions are shown as squares and their color represents the measured thickness. The average thickness for this sample is determined to be $L=(6 \pm 2)\mu\:$m. \textbf{c}, Measured inverse total transmission through the multiple scattering samples versus sample thickness obtained using an integrating sphere. The line represents a linear fit to the data whereby the transport mean free path and the effective refractive index are obtained.}
  \label{fig03}
\end{figure}

The multiple scattering samples are prepared by spreading suspensions of titanium dioxide particles on microscope cover glasses resulting in a typical sample size of $45\,$mm x $25\,$mm. A scanning electron microscope image of a sample is shown in Fig.~\ref{fig03}a. After evaporation of the liquid, the thicknesses of the samples are measured by scanning each sample using the tip of a micrometer screw. The statistics in the thickness determination is obtained by probing 20 different positions on the samples (Fig.~\ref{fig03}b). The central area of each sample is left untouched to avoid damage using this contact technique. Note that all thickness measurements are carried out far from the sample border where surface tension of the suspension will lead to large variations in the sample thickness. The average thicknesses of the samples can be seen in Fig.~\ref{fig03}c, where the error bars display the measured variance in the thickness. In the optical experiments discussed in the next section only a small region in the center of the sample is investigated, and we therefore expect that the recorded size fluctuations provide an upper bound for the uncertainty relevant in the optical experiment.

The ensemble averaged total intensity transmission and reflection coefficients $\overline {T},$ $\overline {R}$ are the characteristic parameters of the multiple scattering medium in the experiments carried out here. Independent measurements of $\overline {T}$ are carried out by using an integrating sphere with two entrance ports with a detector mounted on one port and the samples on the second port. We record the total transmitted power, $P$, through a sample and the light power, $P_0$, without the sample to extract the total sample transmission, $T$, and ensemble average by measuring at six different positions on the sample. Variations in the transmission coefficients were observed depending on which of the two sample surfaces were illuminated, since they are surrounded by different dielectric media (glass and air, respectively). The inverse of the total transmission is in the thin sample approximation given by~\cite{Rivas99}
\begin{equation}\label{eq01_e}
\overline{T}^{-1}=\frac{L+z_{e_1}+z_{e_2}}{\ell+z_{e_1}}\times\frac{1}{T_\text{surf}}.
\end{equation}
 Here $z_{e1}$ and $z_{e2}$ are the extrapolation lengths of the medium that are related to the effective refractive index~\cite{physreva44p3948} and $T_\text{surf}$ accounts for the reflections at the two sample boundaries that depend on the refractive indices of the surrounding materials (either $n_\text{glass} = 1.45$ or $n_\text{air} = 1.0$). Fitting the experimental data of Fig.~\ref{fig03}c with this theory, we estimate a transport mean free path of $\ell = 0.9\pm 0.3\,\mu$m and an effective refractive index of $n_\text{eff} = 2.0 \pm 0.4$ corresponding to extrapolation lengths of $z_{e1} = 4.3\pm0.3\,\mu$m and $z_{e2} = 4.6 \pm 0.3\,\mu$m, respectively. From that the average number of scattering events, $N=(L_\text{eff}/\ell)^2$, taking place in the multiple scattering media can be estimated, where $L_\text{eff}=L+z_{e1}+z_{e2}$ is the effective sample thickness including interface effects.~\cite{prl60p1134} For our samples we estimate $N \approx 300-1000$ scattering events for sample thicknesses varying between $L=6\,\mu$m and $L=20\,\mu$m demonstrating that we are deep in the multiple scattering regime.

\begin{figure}[ht]
  \centering
  \includegraphics[width=0.4\textwidth]{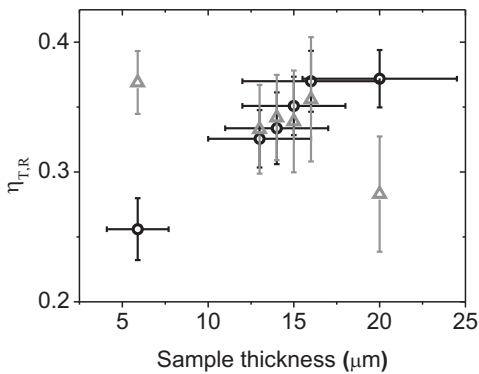}
\caption{Overall collection efficiency for each sample thickness as measured in transmission ($\eta_T$, gray triangles) and reflection ($\eta_R$ black circles). The error bars on the sample thickness are included only on the reflection data.}
  \label{fig04}
\end{figure}

A crucial issue for any quantum optics experiments in multiple scattering media is to collect a sufficiently amount of photons  in order to see nonclassical features. Consequently, the total collection efficiencies in both transmission and reflection $(\eta_{T,R})$ are essential parameters that can be measured independently from the quantum noise measurements by recording the ratio of the total transmitted (reflected) power collected with the squeezing detector relative to the total power transmitted through (reflected from) the sample. In the reflection setup, the multiple scattering medium is illuminated under a steep angle of $69^\circ$. The fraction of the incident light that is multiply scattered is $\overline R=T_\text{surf}(\alpha)\overline{R_s}$. The light that enters the sample, $T_\text{surf}(69^\circ)\approx 72\,\%$, is calculated from Fresnel relations at the sample surface. The amount that is reflected inside the sample, $\overline{R_s}=1-\overline{T_s}$, is obtained from sample transmission measurements with an integrating sphere using $\overline {T} = \overline{T_\text{s}}\,T_\text{surf}(0^\circ)$ and incorporating the surface reflection for light under normal incidence to the multiple scattering medium, see Eq.~\eqref{eq01_e}. The detection efficiency for reflection measurements is then given by $\eta_R=\left(\overline{P_\text{det}}/P_0\right)/\overline R$, where $P_0$ is the power of the light source and $\overline{P_\text{det}}$ denotes the ensemble-averaged power
of the reflected multiply scattered light detected by the squeezing detector. The measured detection efficiencies for the transmission and reflection experiments are presented in Fig.~\ref{fig04}. They are found to vary with the thickness of the sample since the diffuse light exiting the random medium is collected with different efficiencies depending on the extent of the diffusion process. We note that the collection efficiency obtained with the microscope objective is much higher than that of an integrating sphere ($\eta\approx1-10\,\%$), although the numerical aperture (NA) of the objective limits the obtainable collection efficiency.

\subsection{Experimental results}

\begin{figure}[ht]
  \centering
  \includegraphics[width=0.5\textwidth]{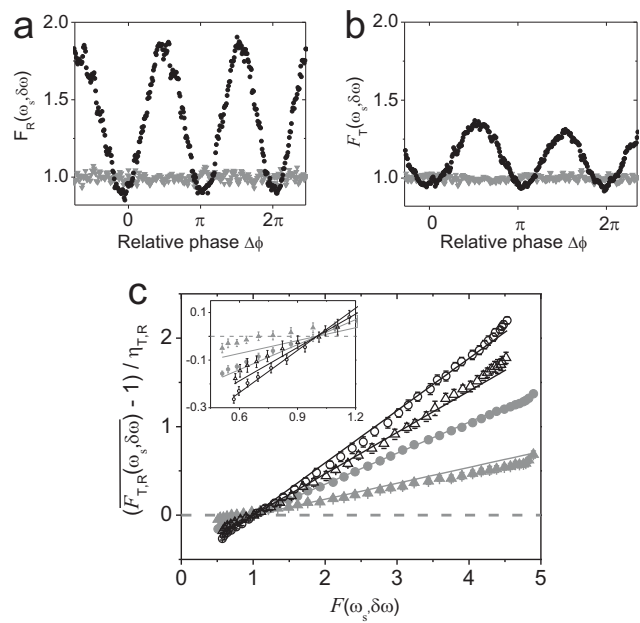}
\caption{Measured Fano factor $F_{T,R}(\omega_s,\delta \omega)$ after multiple scattering of a squeezed state, as recorded in \textbf{a}, reflection geometry and \textbf{b}, transmission geometry for $L=16\,\mu\text{m}$ (black circles). The classical limit is depicted with gray triangles. \textbf{c}, Ensemble-averaged transmitted (filled gray symbols) and reflected (empty black symbols) Fano factor depending on the incident Fano factor, $F$. The inset shows a zoom in of the data that have been taken for Fano factors smaller than $F<1$ highlighting the quantum regime of multiple scattering. The sample thicknesses are $6\,\mu$m (triangles) and $20\,\mu$m (circles), respectively. The theory expressed by Eq. (\ref{eqns:Fano}) is shown by the straight lines and the classical limited is given by the dashed line. The ensemble average has been performed for six different sample positions and the result has been verified 18 times per sample position.}
  \label{fig05}
\end{figure}

The Fano factor of multiply scattered light, measured in reflection geometry and depending on the relative phase $\Delta\phi,$ is displayed in Fig.~\ref{fig05}a for a sample of $L=16\,\mu$m. Using the bright squeezed light source, we observe photon fluctuations of the multiply scattered photons that clearly differ from the classical limit. Either classical or nonclassical photon fluctuations are recorded depending on $\Delta\phi$. For comparison, Fig.~\ref{fig05}b shows the Fano factor of the total transmission. In both cases the measurement is performed at a single sample position, representing a single realization of disorder. The observed variation in the Fano factor of the transmitted light is smaller than in the reflection. This can be explained by a higher reflection than transmission of our sample since the Fano factor is proportional to the sample transmission and reflection, respectively (cf. Eq.~\eqref{eqns:Fano}).

 Multiple scattering is a random process and Fig.~\ref{fig05}a, b are examples of data recovered for a single realization of disorder that consequently are unpredictable since all physical observables in multiple scattering are statistical quantities. In order to compare the experiment with theory, we perform an ensemble average over different realizations of disorder. First, we tune the relative angle $\Delta\phi$ over 18 full periods ($0, 36\pi$) at a single sample position. Data taken at $\Delta\phi+2\pi N$ for different $N$ correspond to the same incident quantum state of light and can be used to average in order to account for fluctuations in the light source. The statistical ensemble average is obtained by repeating the procedure for six different sample positions. Figure~\ref{fig05}c shows the transmitted and reflected, ensemble-averaged Fano factor depending on the incident quantum state of light for two different samples thicknesses. The output Fano factors are found to depend linearly on the Fano factor of the light source, $F(\omega_s,\delta \omega)$, in very good agreement with the quantum theory of multiple scattering, cf. Eq.~\eqref{eqns:Fano}, after including the independently measured collection efficiency (Fig.~\ref{fig04}). For a nonclassical light source ($F(\omega_s,\delta \omega)<1$), the multiply scattered photons have reduced photon fluctuations below the classical limit even after performing the ensemble average demonstrating that nonclassical properties of light can be detected after the random process of multiple scattering (inset Fig.~\ref{fig05}c). This phenomenon that is of purely quantum origin could open new possibilities to increase the information capacity in communication applications since the amount of transmitted information can be increased in a strongly scattering environment.~\cite{science287p287} The classical limit of the information capacity is given by the signal-to-noise ratio of the detected light utilizing a light source with Poissonian photon statistics. Using nonclassical light as in the present work breaks this classical limit and therefore further increases the information capacity.~\cite{prl89p043902}

\begin{figure}[ht]
  \centering
  \includegraphics[width=0.45\textwidth]{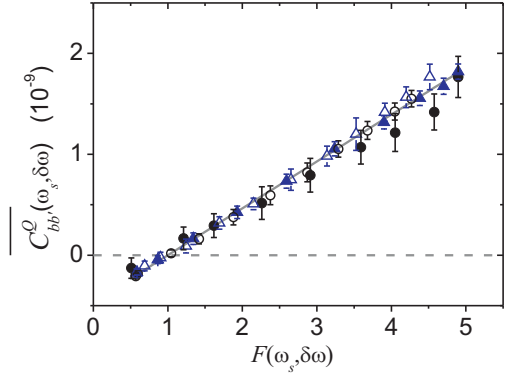}
\caption{(color online). Measured spatial quantum correlation function depending on the incident Fano factor for different sample thicknesses (blue triangles: $6\: \mu$m, black circles: $20\: \mu$m). The spatial quantum correlation function has been measured in transmission (filled symbols) and reflection (empty symbols), respectively. The solid gray line displays the theoretical prediction while the dashed gray line represents the quantum limit of $\overline{C^Q_{bb'}}(\omega_s,\delta \omega)=0$. For visibility we only show selected data points for each measurement.}
  \label{fig06}
\end{figure}

Finally, we study the spatial correlations induced by multiple scattering of squeezed light. $\overline{C^Q_{bb'}}(\omega_s,\delta \omega)$  does not depend on the spatial separation of the modes $b$ and $b'$ in the diffusive regime, i.e. the spatial quantum correlation is of infinite range. Therefore the correlation function can be determined from the photon fluctuations of the total transmittance or reflectance, as predicted in Eq.~\eqref{corr_freq_transmitted}. Thus, we record the ensemble-averaged Fano factor $\overline{F_{T,R}}(\omega_s,\delta \omega)$  and the ensemble-averaged light power of the multiple scattered light. Using the calibrated photo detector and an incident light power of $P=120\,\mu$W and $\hbar \,\omega_0 = 1.17 \: \mbox{eV}$ ($\lambda_0 = 1064 \: \mbox{nm} $) we determine the incident mean photon flux get $\langle \hat n(t) \rangle = P/\hbar \,\omega_0=6.46\times10^{14} \: \mbox{s}^{-1}$.

Figure~\ref{fig06} shows that the spatial quantum correlation function in a multiple scattering medium can be controlled by the quantum state of the incident light source. The correlation function is obtained from quantum noise measurements of the transmitted and reflected light, as outlined in Sec.~\Rmnum{2}B. The data are shown for two different samples ($L=6\,\mu$m and $L=20\,\mu$m) as a function of the incident Fano factor, $F(\omega_s,\delta \omega)$. We clearly observe a linear dependence of $\overline{C^Q_{bb'}}(\omega_s,\delta \omega)$ on the Fano factor of the light entering the medium. The incident Fano factor and photon flux were measured by removing the sample. In excellent agreement with theory (Eq.~\eqref{corr_freq_incident}), we find in transmission and reflection measurements that $\overline{C^Q_{bb'}}(\omega_s,\delta \omega)$ is independent of the sample thickness. The strength of the spatial quantum correlation varies over a broad range between $-2\times10^{-10}$ and $1.8\times10^{-9}$. Note that in Ref.~\onlinecite{prl102p193901} the bandwidth of the measurement was not accounted for properly, which underestimated the magnitude of the spatial correlation function \cite{erratum}. We highlight that our observation of non-vanishing spatial quantum correlations that are tunable above and below the classical limit is in stark contrast to the classical intensity correlation function that is always unity in the diffusive regime.~\cite{physreve74p045603} The experiment demonstrates that the nonclassical photon statistics can survive  multiple scattering and persists even after ensemble averaging. An interesting next step will be to study experimentally the quantum interference of two or more independent quantum states in a multiple scattering medium.~\cite{prl105p090501}

\section{Conclusion}

In conclusion, we have presented a continuous-mode quantum theory of multiple scattering in order to properly describe spatial quantum correlations that are induced by multiple scattering. Depending on the quantum state of the incident light source we predict and observe that the multiple scattered light can exhibit either classical or nonclassical photon fluctuations. From total transmission and total reflection measurements we observed either classical or nonclassical spatial correlations when varying the quantum state of the squeezed light source. Controlling the quantum state of the incident light source provides an efficient way of tuning continuously the multiple scattering spatial quantum correlations. The experimental results were in excellent agreement with the theoretical analysis. We expect our results might inspire new experiments on, e.g., multiple scattering of entangled photons or quantum interference.~\cite{prl102p193601,prl104p173601,prl105p090501,prl105p163905}

\section{Acknowledgements}
We thank Jir\'i Janousek for help with the nonclassical light
source, Elbert G. van Putten, Ivo M. Vellekoop, and Allard P.
Mosk for providing the samples, and Ad Lagendijk for stimulating discussions. We gratefully acknowledge the
Danish Council for Independent Research (Technology and Production Sciences (FTP) and Natural Sciences (FNU)) and the European Research Council (ERC Consolidator Grant "ALLQUANTUM" for P.L.) for financial support. A.H. and U.L.A. acknowledge support from the EU project COMPAS.

\newpage

\end{document}